\documentclass[onecolumn]{IEEEtran}
\usepackage{amsmath,bm}
\usepackage{amsfonts}
\usepackage{amssymb}
\usepackage[noadjust]{cite}
\ifCLASSINFOpdf
   \usepackage[pdftex]{graphicx}
   \DeclareGraphicsExtensions{.pdf,.jpeg,.png}
\else
   \usepackage[dvips]{graphicx}
   \DeclareGraphicsExtensions{.eps}
\fi

\begin{document}

\title{Coded DS-CDMA Systems with Iterative Channel Estimation and no Pilot
Symbols}
\author{Don~Torrieri,~\IEEEmembership{Senior Member, IEEE,} Amitav~Mukherjee,~\IEEEmembership{Student~Member, IEEE,}
and~Hyuck~M.~Kwon,~\IEEEmembership{Senior~Member,~IEEE} \thanks{%
Don Torrieri is with the US Army Research Laboratory, Adelphi, MD 20873 USA
(email: dtorr@arl.army.mil).} \thanks{Amitav Mukherjee is with the
Department of Electrical Engineering and Computer Science, University of
California, Irvine, CA 92617 USA (email: a.mukherjee@uci.edu).} \thanks{%
Hyuck M. Kwon is with the Department of Electrical Engineering and Computer
Science, Wichita State University, Wichita, KS 67260 USA (e-mail:
hyuck.kwon@wichita.edu).}\thanks{%
This work was partly sponsored by the Army Research Office under DEPSCoR ARO
Grant W911NF-08-1-0256, and by NASA under EPSCoR CAN Grant NNX08AV84A.}}


\maketitle

\begin{abstract}
In this paper, we describe direct-sequence code-division multiple-access
(DS-CDMA) systems with quadriphase-shift keying in which channel estimation,
coherent demodulation, and decoding are iteratively performed without the
use of any training or pilot symbols. An expectation-maximization
channel-estimation algorithm for the fading amplitude, phase, and the
interference power spectral density (PSD) due to the combined interference
and thermal noise is proposed for DS-CDMA systems with irregular
repeat-accumulate codes. After initial estimates of the fading amplitude,
phase, and interference PSD are obtained from the received symbols,
subsequent values of these parameters are iteratively updated by using the
soft feedback from the channel decoder. The updated estimates are combined
with the received symbols and iteratively passed to the decoder. The
elimination of pilot symbols simplifies the system design and allows either
an enhanced information throughput, an improved bit error rate, or greater
spectral efficiency. The interference-PSD estimation enables DS-CDMA systems
to significantly suppress interference.
\end{abstract}







\section{INTRODUCTION}

In mobile communication systems, the wireless channel induces random
amplitude and phase variations in the received data, with the possible
addition of time-varying interference from co-channel users. For this
reason, the accuracy of channel state information (CSI) at the receiver is
critical for coherent detection and demodulation. A number of methods have
been proposed for estimation of CSI, all of which fall within the broad
categories of either pilot-assisted or blind algorithms. Current and
next-generation cellular protocols such as W-CDMA (Wideband Code Division
Multiple Access) and 3GPP LTE (Third Generation Partnership Project
Long-Term Evolution) specify the use of pilot-assisted channel estimation
(PACE) \cite{3GPP}. Pilot symbols or training sequences are known symbols
either multiplexed with or superimposed onto the transmitted data in the
time or frequency domain, with the associated disadvantage of a loss in
spectral and/or power efficiency. Moreover, superimposed PACE is degraded at
low signal-to-noise ratios, and multiplexed PACE is unsuitable for
fast-fading channels with a coherence time shorter than the pilot-symbol
transmission rate \cite{Cavers}, \cite{Globe99}.

\emph{Blind channel-estimation methods} offer an alternative approach that
avoids the implementation cost of pilot symbols \cite{Blind94}. Blind
methods typically use second-order statistics of the received symbols for CSI estimation, with shortcomings
such as increased complexity, slow convergence times, and channel-phase
ambiguity \cite{Poor98}. In addition, the received \textit{interference
power spectral density} (PSD), which is due to both the thermal noise and
the time-varying interference, is usually not estimated in the literature
spanning both PACE and blind CSI estimation. The accuracy of the
interference-PSD estimation is known to have a significant impact on
turbo-principle (iterative) detection techniques as well as turbo and
low-density parity-check (LDPC) channel decoding \cite{SNR}, \cite%
{Mackay_noise}.

The expectation-maximization (EM) algorithm offers a low-complexity
iterative approach to optimal maximum-likelihood detection and
estimation \cite{EM}, \cite{EM_1977}. A substantial body of literature can be
found on EM-based techniques for data detection, multiuser detection,
channel estimation, or a combination of the latter. A few representative
examples are listed next. A recursive estimation of the fading channel
amplitude was proposed in \cite{EM_99}. Iterative receivers with EM-based
fading-amplitude and data estimation using pilot symbols for LDPC-based
space-time coding and space-time block-coded orthogonal frequency-division
multiplexing (OFDM) were studied in \cite{EM_Wang1} and \cite{EM_Wang2},
respectively. Joint multiuser detection and channel/data estimation for
uplink code-division multiple access (CDMA) was studied in \cite{EM_ref}--%
\cite{EM_Mitra}. In \cite{ChengTrans07}, iterative EM estimation and turbo
coding were studied assuming noncoherent frequency-shift keying modulation
and demodulation, which is well-known to be less power-efficient than
coherent modulation \cite{Proakis}.

In \cite{MILCOM06}, an EM estimation approach for turbo-coded single-user
iterative CDMA receivers with binary phase-shift keying was considered. In
\cite{VTC07} and \cite{MILCOM07}, the authors replaced turbo codes with
regular LDPC codes; however, \cite{MILCOM06}--\cite{MILCOM07} all featured
as much as a 9.1\% pilot-symbol overhead for channel-amplitude and
interference-PSD estimation. Recently, EM-based channel and noise estimation
techniques were proposed in \cite{Vandendorpe2007} and \cite{Choi} for
multiple-antenna systems with convolutional coding and as much as a 10\%
pilot-symbol overhead for initial channel estimation.

Although the primary role of pilot symbols in most cellular standards is
channel estimation, pilot symbols often play a secondary role in cell,
frame, or symbol synchronization. However, alternative methods of
synchronization may be used when pilot symbols are unavailable \cite{Proakis}, \cite{Psar1},
\cite{Psar2}. In this paper, a doubly iterative direct-sequence CDMA (DS-CDMA) receiver
featuring iterative EM channel estimation and iterative detection and
decoding without \emph{any} pilot symbols is presented. The general form of
the proposed blind channel estimator provides fading-amplitude, phase, and
interference-PSD estimates in both single-user and multiuser environments,
therefore offering an alternative to the methods proposed in \cite{Stuber}
and \cite{Pados07} to rectify the phase ambiguity of blind channel estimates%
\footnote{%
In \cite{Stuber}, two different PSK modulations are used on adjacent OFDM
subcarriers to resolve the phase ambiguity under slow frequency-selective
fading. A short pilot sequence is used in \cite{Pados07} to recover the
channel phase, making it semi-blind in nature. More importantly, the
interference-plus-noise PSD is not estimated in \cite{Stuber} and \cite{Pados07}.}. The special case of EM channel estimation with perfect phase information
at the receiver (e.g., by means of a phase-locked loop) is also considered.
The proposed iterative receiver is capable of using higher-order modulations
such as M-PSK and M-ary quadrature amplitude modulation (M-QAM), although
quadriphase-shift keying (QPSK) is demonstrated in this work for simplicity.
In addition, the proposed system uses irregular repeat-accumulate (IRA)
codes instead of regular LDPC codes for lower complexity \cite{Yang04}--\cite{Richardson01}.

The paper is organized as follows. Section II describes the system
transmitter and receiver models including coding, modulation, and spreading,
as well as fading-channel parameters. Section III summarizes the proposed
EM-based estimation process that uses soft feedback from the channel
decoder. Section IV presents the proposed blind method for the initial CSI
estimation and the possible trade-offs vis-\`{a}-vis PACE. Section V shows
simulation results, and Section VI offers conclusions.

A word on notation: lowercase boldface is used to represent vectors, while
uppercase boldface represent matrices. $E$ denotes the statistical
expectation, $\left( \cdot\right) ^{T}$ is the matrix transpose, * is the
complex conjugate, and $\lfloor x\rfloor$ is the largest integer smaller
than $x$.

\section{SYSTEM MODEL}
\begin{figure}[htbp]
\centering
\includegraphics[width=\linewidth]{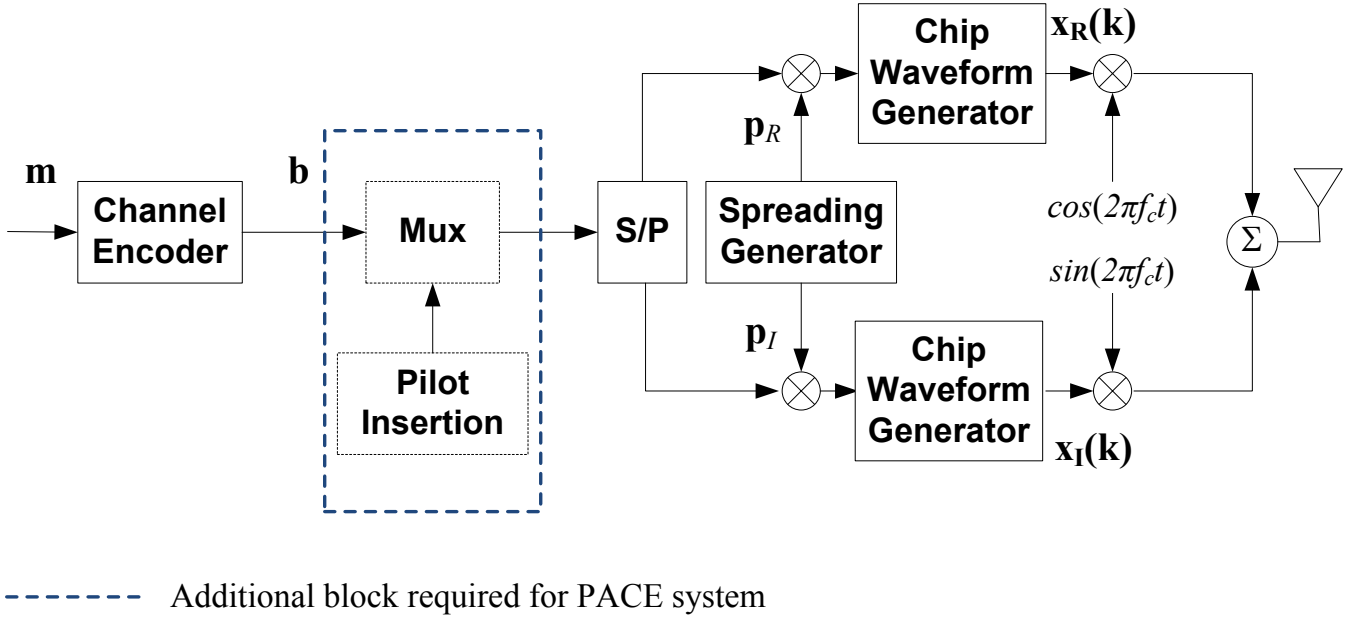}
\caption{DS-CDMA transmitter with QPSK modulation.}
\label{fig:TX}
\end{figure}
Fig. 1 shows the block diagram of a dual quaternary DS-CDMA transmitter \cite%
{Torrieri} consisting of a channel encoder, QPSK modulator, and a
direct-sequence spreading generator that multiplies orthogonal chip
sequences ${\mathbf{p}}_{R}$ and ${\mathbf{p}}_{I}$ with the in-phase and
quadrature modulator inputs. The input to the encoder in Fig. 1 is a binary,
independent, identically distributed data block of length $K$, which is
denoted by $\mathbf{m}=\left[ m(1),\ldots,m(K)\right] $, $%
m(i_{bit})\in\lbrack1,0].$

\subsection{Encoding, Modulation, and Spreading}

Each $1\times K$ message vector $\mathbf{m}$ is encoded into a $1\times N$
codeword $\mathbf{b}=[b(1),\ldots,b(N)]$ using a systematic, extended IRA
code \cite{Yang04}. IRA codes offer a combination of the linear complexity
of turbo encoding and the lower complexity of LDPC decoding without
compromising on performance.

The $\left( {N,K}\right)$ IRA code is constructed following the
methodology proposed in \cite{Ryan06}, where the IRA code parameters were
designed for use on a burst-erasure channel with additive noise, which was
shown to be a good surrogate for Rayleigh fading channels. IRA codes can be
considered to be a subset of low-density parity-check codes and therefore
may be represented by a Tanner graph \cite{Richardson01}. Let $\lambda\left(
x\right) =\sum\nolimits_{i}^{d_{v}}{\lambda_{i}}x^{i-1}$ and $\rho\left(
x\right) =\sum\nolimits_{i}^{d_{c}}{\rho_{i}}x^{i-1}$ represent the variable-node and
check-node degree distributions of the code's Tanner graph, with $%
\left( d_{v},d_{c}\right) $ being the maximum variable and check node
degrees, respectively. Using density evolution, for $\left(
d_{v}=8,d_{c}=7\right) $ we obtain the following good choices \cite{Ryan06}:
\begin{eqnarray}
\lambda\left( x\right) &=&0.00008+0.31522x+0.34085x^{2}+0.0.06126x^{6}\nonumber \\
                         &&{+}\: 0.28258x^{7} \nonumber\\
\rho\left( x\right) &=&0.62302x^{5}+0.37698x^{6}.  \label{EQ:degree}
\end{eqnarray}
The $\left( {N-K}\right) \times N$ IRA parity-check matrix can be
represented as ${\mathbf{H}}=\left[ {{\mathbf{H}}_{1}{\text{ }}\mid{\text{ }
}{\mathbf{H}}_{2}}\right] $, where sub-matrix $\mathbf{H}_{2}$ is a $\left( {%
\ N-K}\right) \times\left( {N-K}\right) $ dual-diagonal matrix, and $\mathbf{%
H }_{1}$ is a randomly-generated $\left( {N-K}\right) \times K$ sparse
matrix constructed such that $\mathbf{H}$ has the degree profile of (\ref%
{EQ:degree}). The $K\times N$ systematic generator matrix $\mathbf{G}$ is
then given by $\mathbf{G}=\left[ {{\mathbf{I}_{K}}{\text{ }}\mid{\text{ }}{%
\mathbf{H}} _{1}^{T}{\mathbf{H}}_{2}^{-T}}\right] $.

For the simulations in Section V, Gray-labeled QPSK is used with 2 encoded
bits mapped into a modulation symbol $x(k)\in \left\{ {\pm 1,\pm j}\right\}
,k=1,\ldots ,\frac{N}{2}$. Although QPSK is assumed, the analysis and
simulation is easily extended to M-QAM. Parallel streams of code bits are
each spread using a Gold sequence with spreading factor $g$ chips/code bit
before rectangular pulse-shaping that produces the real and imaginary
components of $x(k)$, i.e., ${x_{R}(k)=\operatorname{Re}\left( {x\left( k\right) }%
\right) }$ and ${x_{I}\left( k\right) =\operatorname{Im}\left( {x\left( k\right) }%
\right) }.$ In practice, an intermediate frequency is used before the
carrier frequency upconversion, but the upconversion from baseband to the
intermediate frequency is omitted for clarity in Fig. 1.

No channel interleaving is applied to the IRA code due to the inherent
interleaving characteristics of the IRA code itself. This is because the IRA
code can be alternatively represented as a repetition code concatenated with
a convolutional encoder (accumulator) with an interleaver between them. The
interleaver is embedded within the sub-matrix $\mathbf{H}_{1}$ in the Tanner
graph representation of IRA codes.

\subsection{Channel Model}

For multiple-access interference (MAI) environments, the channel
coefficients are generated using the Jakes correlated fading model. The
flat-fading assumption is valid when the information bit-rate is low, e.g.,
100 kb/s as usually considered in this paper, since the multipath delay
spread in a typical cellular environment is about 10 $\mu s,$ which is
negligible compared to the symbol duration. For completeness, the proposed
system and analysis are extended to include frequency-selective channels by
including multipath components with delays exceeding a chip duration and
using Rake receivers \cite{Proakis}, \cite{Torrieri}, as described in
Section~\ref{sec:RAKE}. Each codeword or frame of $N$ code bits is divided
into two different types of subframes or blocks. One block size is set equal
to the $n_{FB}$ code bits over which the fading amplitude is assumed to be
constant. The other block size is set equal to $n_{IB}$ code bits over which
the interference level is assumed to be constant.

Each frame comprises $N/2$ QPSK code symbols and $Ng/2$ spreading-sequence
chips for each QPSK component. The fading coefficient associated with
spreading-sequence chip $c$ of either ${\mathbf{p}}_{R}$ or ${\mathbf{p}}%
_{I} $ is
\begin{equation*}
C_{_{\left\lfloor {c/(n_{FB}g)}\right\rfloor }}=\sqrt{E_{s}}\alpha
_{\left\lfloor {c/(n_{FB}g)}\right\rfloor }e^{j\phi_{\left\lfloor {%
c/(n_{FB}g)}\right\rfloor }},
\end{equation*}%
\begin{equation}
c=1,\ldots,\frac{Ng}{2}
\end{equation}
where $E_{s}$ is the average energy per QPSK symbol, $\alpha$ is the fading
amplitude with $E\left[ {\alpha^{2}}\right] =1$, and $\phi$ is the unknown
fading-induced channel phase.

\subsection{Iterative Receiver Structure}
\begin{figure}[htbp]
\centering
\includegraphics[width=\linewidth]{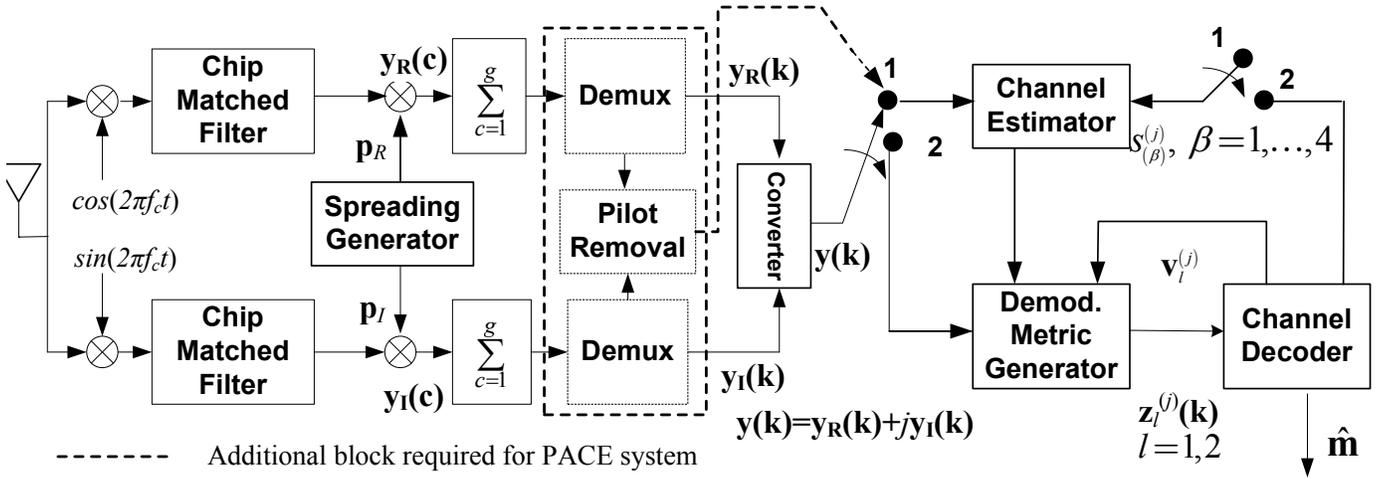}
\caption{Iterative DS-CDMA receiver.}
\label{fig:RX}
\end{figure}
Fig. 2 shows a block diagram of the proposed dual quaternary iterative
receiver. The received signal is downconverted, passed through
chip-matched filters, and despread by a synchronized spreading-sequence
generator in each branch, with the downconverter and synchronization devices
\cite{Proakis} omitted in Fig. 2 for clarity. Self-interference between the
spreading sequences of the desired user is negligible because accurate
synchronization is assumed at the receiver. Let $N_{0}/2$ denote the
two-sided PSD of the Gaussian noise. For the flat-fading scenario, the
complex envelope of the desired user at the $k^{th}$ symbol time with active
MAI can be written as
\begin{equation}
y\left( k\right) =C_{_{\left\lfloor {k/n_{FB}}\right\rfloor
}}x(k)+n^{int}(k)+n(k),\text{ }1\leq k\leq \frac{N}{2}
\end{equation}%
where $x(k)$ is the complex transmitted code symbol of the desired user, $%
n(k)$ is a complex zero-mean circularly symmetric Gaussian noise sample with
$E\left[ {\left\vert {n_{k}}\right\vert ^{2}}\right] =N_{0}$, and $%
n^{int}(k) $ is the interference at the demodulator due to interfering users
\cite{Proakis}, \cite{Torrieri}.

The time-varying MAI is assumed to be generated by interfering users with a
structure identical to the desired user, albeit the spreading sequences
differ and the fading coefficients are independent. The despreading in the
receiver tends to whiten the interference PSD over the code-symbol passband,
and the subsequent filtering tends to produce a residual interference with a
Gaussian distribution. Thus, the \textit{ interference PSD} due to the
combined interference and thermal noise is modeled as additive Gaussian
noise with a two-sided PSD $I_{0}/2$ that is constant over each block of $%
n_{IB}$ code bits but varies from block-to-block. This model enables the
derivation of an EM estimator for $I_{0}$ that is used in the demodulator
metric and leads to the suppression of the interference.

A \emph{receiver iteration} is defined as a fixed number of decoder
iterations followed by internal EM iterations in the channel estimator of
Fig. 2, and then a single demodulator metric generation. Let

\begin{description}
\item[$i$] denote the index for the internal EM iteration , $i=1,\ldots
,i_{max}$;

\item[$j$] denote the index for the closed-loop receiver iteration, $%
j=1,\ldots, j_{max}$.
\end{description}

Let $\bm{\hat{\theta}}_{(i)}^{(j)}=\left( {\hat{C}_{(i)}^{(j)},\hat{I}%
_{0(i)}^{(j)}}\right) $ represent the estimates of the fading-coefficient
and interference-PSD parameters at the $i^{th}$ EM iteration during the $%
j^{th}$ overall receiver iteration. EM iterations commence after the initial
channel estimation and decoding, which is obtained while the switch in Fig. %
2 is set to position 1. The subsequent receiver iterations are
performed while the switch is set to position 2 in order to refine the
initial channel estimate with the aid of soft feedback $s_{\beta }^{(j)}$, $%
\beta =1,2,3,4$ from the channel decoder.

\section{EM ALGORITHM}

Theoretically, the maximum-likelihood CSI estimator $\bm{\hat{\theta}}$ can be
obtained from a received data vector ${\mathbf{y}}=[y(1),\ldots,y(N_{1})]$
of $N_{1}$ code symbols, referred to as the \emph{incomplete data}, by
maximizing the conditional log-likelihood function:
\begin{equation}
\bm{\hat{\theta}}=\arg\mathop {\max }\limits_{\bm{\theta}}\ln f({\mathbf{y}}%
\mid\bm{\theta}).
\end{equation}
However, the computation of this equation is virtually prohibitive in
practice since its complexity increases exponentially with the observation
window size. In the EM algorithm, the expectation of the conditional
log-likelihood of the \emph{complete data} $\mathbf{z}=(\mathbf{x},\mathbf{y}%
)$ is iteratively maximized with respect to $\bm\theta$, where expectation is
taken with respect to $\mathbf{x}$ given $\mathbf{y}$ and a previous
estimate of $\bm{\theta}$.

The conditional probability density function (pdf) of $\mathbf{z}$ can be
written as
\begin{equation}
f(\mathbf{z}\mid \bm\theta )=f(\mathbf{x},\mathbf{y}\mid \bm\theta )=f(\mathbf{y}%
\mid \mathbf{x},\bm\theta )f(\mathbf{x}\mid \bm{\theta} )=f(\mathbf{y}\mid \mathbf{x}%
,\bm\theta )f(\mathbf{x})
\end{equation}%
where the last equality is from the independence of the transmitted signal
vector $\mathbf{x}$ and the CSI parameter $\bm\theta $. Thus,
\begin{equation}
\ln f\left( {\mathbf{z}\mid \bm\theta }\right) =\ln f\left( {\mathbf{y}\mid
\mathbf{x},\bm\theta }\right) +\ln f\left( \mathbf{x}\right) .  \label{EQ:EM1}
\end{equation}%
Since the symbols are independent and circularly symmetric Gaussian noise
and interference are assumed, the conditional pdf $f\left( {\mathbf{y}\mid
\mathbf{x},\bm\theta }\right) $ is
\begin{equation}
f\left( {\mathbf{y}\mid \mathbf{x},\bm\theta }\right) =\frac{1}{{(\pi
I_{0})^{N_{1}}}}\exp \left( {-\sum\limits_{k=1}^{N_{1}}{\frac{{\left( {%
|y(k)-Cx(k)|^{2}}\right) }}{{I_{0}}}}}\right) .  \label{EQ:EM1a}
\end{equation}%
Therefore, as $\left\vert {x(k)}\right\vert ^{2}=1\hspace{0.13in}\forall k,$
\begin{eqnarray}
\ln f\left( {\mathbf{y}\mid \mathbf{x},\bm\theta }\right) &=&-N_{1}\cdot \ln
\left( {I_{0}}\right) -\frac{1}{{I_{0}}}\sum\limits_{k=1}^{N_{1}}[ %
\left\vert {y(k)}\right\vert ^{2}+|C|^{2}\nonumber\\
&&{-}\: 2\operatorname{Re}\left( {y^{\ast }(k)Cx(k)%
}\right) ]   \label{EQ:EM2}
\end{eqnarray}%
where an irrelevant constant has been dropped.

\textbf{E-step}: Define the objective function to be the conditional
expectation of the conditional log-likelihood of $\mathbf{z}=(\mathbf{y},%
\mathbf{x})$, which can be written as
\begin{equation}
\chi\left( {\bm{\theta}},{\bm{\hat{\theta}}}_{(i)}^{(j)}\right) =E_{{%
\mathbf{z}}\mid{\mathbf{y}},{\bm{\hat{\theta}}}_{(i)}^{(j)}}\left[ {\ln
f({\mathbf{z}}\mid{\bm{\theta}})}\right]  \label{EQ:Obj}
\end{equation}
where ${\bm{\hat{\theta}}}_{(i)}^{(j)}$ is the previous estimate. Using (%
\ref{EQ:EM1}) and (\ref{EQ:EM2}) and observing that $\ln f\left( \mathbf{x}%
\right) $ in (\ref{EQ:EM1}) is independent of $\bm\theta$, and hence irrelevant
to the maximization, we obtain
\begin{eqnarray}\label{EQ:objective}
\chi\left( \bm\theta,\bm{\hat{\theta}}_{(i)}^{(j)}\right) &=&-N/2\cdot\ln\left( {I_{0}}%
\right) -\frac{1}{{I_{0}}}\sum\limits_{k=1}^{N_{1}} [\left\vert {y(k)}%
\right\vert ^{2}+|C|^{2}\nonumber\\
&&{-}\: 2\operatorname{Re}\left( {y^{\ast }(k)C\bar{x}
_{(i)}^{(j)}(k)}\right)]
\end{eqnarray}
where $\bar{x}_{(i)}^{(j)}(k)=E_{\mathbf{z}\mid{{\mathbf{y}},\bm{\hat{\theta }}%
_{\left( {i}\right) }^{(j)}}}\left[ {x(k)}\right] =E_{\mathbf{x}\mid\mathbf{y%
},\bm{\hat{\theta}}_{\left( {i}\right) }^{(j)}}\left[ {x(k)}\right] .$ Assuming
the independence of each transmitted symbol $x(k)$ and the independence of $%
x(k)$ and $\bm{\hat{\theta}}_{\left( {i}\right) }^{(j)},$ and using Bayes' law
and the fact that (\ref{EQ:EM1a}) can be expressed as a product of $N_{1}$
factors, we obtain
\begin{equation}
\bar{x}_{(i)}^{(j)}(k)=E_{x(k)\mid{y(k)},\bm{\hat{\theta}}_{\left( {i}\right)
}^{(j)}}\left[ {x(k)}\right]
\end{equation}
where
\begin{equation}
f\left( {x(k)\mid y(k),{\bm{\hat{\theta}}}_{(i)}^{(j)}}\right) =\frac{{%
f\left( {y(k)\mid x(k),{\bm{\hat{\theta}}}_{(i)}^{(j)}}\right) }}{{%
f\left( {y(k)\mid{\bm{\hat{\theta}}}_{(i)}^{(j)}}\right) }}\Pr\left( {%
x(k)}\right) .  \label{EQ:EM3}
\end{equation}
and
\begin{equation}
{f\left( {y(k)\mid x(k),{\bm{\hat{\theta}}}_{(i)}^{(j)}}\right) =}\frac{1%
}{{\pi I_{0}}}\exp\left( {-{\frac{{{|y(k)-Cx(k)|^{2}}}}{{I_{0}}}}}\right) .
\end{equation}

\textbf{M-step}: Taking the derivative of (\ref{EQ:objective}) with respect
to the real and imaginary parts of the complex-valued $C$, and then setting
the results equal to zero, we obtain the estimate of the fading coefficient
at iteration $i+1$ as
\begin{align}
\operatorname{Re}\left( {\hat{C}_{\left( {i+1}\right) }^{\left( j\right) }}\right) &
=\frac{1}{N_{1}}\sum\limits_{k=1}^{N_{1}}{\operatorname{Re}\left( {y^{\ast }\left(
k\right) \bar{x}_{(i)}^{(j)}(k)}\right) }  \label{EQ:ChatRe} \\
\operatorname{Im}\left( {\hat{C}_{\left( {i+1}\right) }^{\left( j\right) }}\right) &
=-\frac{1}{N_{1}}\sum\limits_{k=1}^{N_{1}}{\operatorname{Im}\left( {y^{\ast }\left(
k\right) \bar{x}_{(i)}^{(j)}(k)}\right) }.  \label{EQ:ChatIm}
\end{align}%
Similarly, maximizing (\ref{EQ:objective}) with respect to the interference
PSD $I_{0}$ leads to
\begin{equation}
\hat{I}_{0,\left( {i+1}\right) }^{(j)}=\frac{1}{N_{1}}\sum%
\limits_{k=1}^{N_{1}}{\left\vert {y(k)-\hat{C}_{\left( {i+1}\right) }^{(j)}%
\bar{x}_{(i)}^{(j)}(k)}\right\vert }^{2}.  \label{EQ:I0}
\end{equation}%
The fading phase and amplitude can be explicitly estimated from (\ref%
{EQ:ChatRe}) and (\ref{EQ:ChatIm}), but that is unnecessary.

Let $s_{\beta}^{(j)}$, $\beta=1,2,3,4$, be the code-symbol probabilities
obtained from the soft outputs of the channel decoder, with $s_{1}=\text{Pr}%
\left( x(k)=+1\right) ,s_{2}=\text{Pr}\left( x(k)=+j\right) ,s_{3}=\text{Pr}%
\left( x(k)=-1\right) ,s_{4}=\text{Pr}\left( x(k)=-j\right) $. From (\ref%
{EQ:EM1a}) and (\ref{EQ:EM3}), the expectation of $x(k)$ at the $i^{th}$ EM
and $j^{th}$ receiver iteration is
\begin{equation}
\bar{x}_{(i)}^{(j)}(k)=\frac{{%
s_{1}^{(j)}R_{1,(i)}^{(j)}+js_{2}^{(j)}R_{2,(i)}^{(j)}-s_{3}^{(j)}R_{3,(i)}^{(j)}-js_{4}^{(j)}R_{4,(i)}^{(j)}%
}}{{\sum\limits_{\beta=1}^{4}{s_{\beta}^{(j)}R_{\beta,i}^{(j)}}}}
\end{equation}
where likelihood-ratio $R_{\beta,(i)}^{(j)}$ depends on the current CSI
estimates as
\begin{eqnarray}
R_{1,(i)}^{(j)}&=&\exp\left[ {\frac{{2}}{\hat{I}_{0,{(i)}}^{(j)}}\operatorname{Re}(%
\hat{C}_{{(i)}}^{(j)}y(k))}\right]\nonumber\\
R_{2,(i)}^{(j)}&=&\exp\left[ {\frac{{2}}{%
\hat{I}_{0,{(i)}}^{(j)}}\operatorname{Im}(\hat {C}_{{(i)}}^{(j)}y(k))}\right]\nonumber\\
R_{3,(i)}^{(j)}&=&\exp\left[ {-\frac{{2}}{\hat{I}_{0,{(i)}}^{(j)}}\operatorname{Re}(%
\hat{C}_{{(i)}}^{(j)}y(k))}\right]\nonumber\\
R_{4,(i)}^{(j)}&=&\exp\left[ {-\frac{{2}}{%
\hat{I}_{0,{(i})}^{(j)}}\operatorname{Im}(\hat {C}_{{(i)}}^{(j)}y(k))}\right].
\end{eqnarray}
Therefore, for a given receiver iteration, $\bar{x}_{(i)}^{(j)}(k)$ and $%
R_{\beta ,i}^{(j)}$ are updated $i_{max}$ number of times using decoder
feedback $s_{\beta }^{(j)}$. In the next receiver iteration, after channel
re-estimation, the fading-coefficient and interference-PSD estimates are
updated, and then used at the demodulator and channel decoder to recompute $%
\bar{x}_{(i)}^{(j+1)}(k)$ and $R_{\beta ,(i)}^{(j+1)}$. This process is
repeated again for $i_{max}$ EM iterations, and the aforementioned cycles
continue likewise for subsequent receiver iterations.

In estimating the fading parameters, we set $N_{1}=n_{FB}/2;$ in estimating $%
I_{0}$, we choose $n_{IB}\leq n_{FB}$ and set $N_{1}=$ $n_{IB}/2.$ The EM
estimator first finds the value of ${\hat{C}_{\left( {i}\right) }^{\left(
j\right) }}$ for a fading block of size $n_{FB}.$ Then it finds the value of
$\hat{I}_{0,\left( {i}\right) }^{(j)}$ for each smaller or equal
interference block of size $n_{IB}$ using the value of ${\hat{C}_{\left( {i}%
\right) }^{\left( j\right) }}$ found for the larger or equal fading block.
When pilot symbols are used, we set ${\bar{x}_{(i)}^{(j)}(k)=x(k)}$ for each
known pilot bit, and there are no EM iterations if only known pilot bits are
processed in calculating the channel estimates. The application of the EM
algorithm is to obtain both channel-coefficient and interference-PSD
estimates, which differs from \cite{EM_Wang1}--\cite{EM_ref} where the
emphasis is on data detection, and noise statistics are assumed to be
perfectly known.

Let $l=1,2$ denote the two bits of a QPSK symbol, and $v_{1},v_{2}$ denote
the corresponding log-likelihood ratios that are fed back by the channel
decoder. From \cite{MILCOM07} and \cite[Eqn. 6]{Valenti}, the demodulation
metrics (extrinsic information) $z_{l}^{(j)}(k),l=1,2$ for bits $1,2$ of
symbol $k$ that are applied to the channel decoder are shown at the top of the next page.
\begin{figure*}
\begin{equation}
z_{1}^{(j)}(k)=\log\frac{{\exp\left[ {\frac{2}{{\hat{I}_{0,\left( {i_{\max}}%
\right) }^{\left( j\right) }}}\operatorname{Im}\left( \hat{C}_{(i_{\max
})}^{(j)}y^{\ast}(k)\right) }\right] +\exp\left[ -{\frac{2}{{\hat {I}%
_{0,\left( {i_{\max}}\right) }^{\left( j\right) }}}\operatorname{Re}\left( \hat{C}%
_{(i_{\max})}^{(j)}y^{\ast}(k)\right) +v_{2}}\right] }}{{\exp\left[ {\frac{2%
}{{\hat{I}_{0,\left( {i_{\max}}\right) }^{\left( j\right) }}}\operatorname{Re}\left(
\hat{C}_{(i_{\max})}^{(j)}y^{\ast }(k)\right) }\right] +\exp\left[ -{\frac{2%
}{{\hat{I}_{0,\left( {i_{\max}}\right) }^{\left( j\right) }}}\operatorname{Im}\left(
\hat{C}_{(i_{\max })}^{(j)}y^{\ast}(k)\right) +v_{2}}\right] }}
\label{EQ:extrins1}
\end{equation}
\begin{equation}
z_{2}^{(j)}(k)=\log\frac{{\exp\left[ -{\frac{2}{{\hat{I}_{0,\left( {i_{\max }%
}\right) }^{\left( j\right) }}}\operatorname{Im}\left( \hat{C}_{(i_{\max
})}^{(j)}y^{\ast}(k)\right) }\right] +\exp\left[ -{\frac{2}{{\hat {I}%
_{0,\left( {i_{\max}}\right) }^{\left( j\right) }}}\operatorname{Re}\left( \hat{C}%
_{(i_{\max})}^{(j)}y^{\ast}(k)\right) +v_{1}}\right] }}{{\exp\left[ {\frac{2%
}{{\hat{I}_{0,\left( {i_{\max}}\right) }^{\left( j\right) }}}\operatorname{Re}\left(
\hat{C}_{(i_{\max})}^{(j)}y^{\ast }(k)\right) }\right] +\exp\left[ {\frac{2}{%
{\hat{I}_{0,\left( {i_{\max}}\right) }^{\left( j\right) }}}\operatorname{Im}\left(
\hat{C}_{(i_{\max })}^{(j)}y^{\ast}(k)\right) +v_{1}}\right] }}.
\label{EQ:extrins2}
\end{equation}
\hrulefill
\end{figure*}

The number of EM iterations and the receiver latency are reduced by applying
a \textit{stopping criterion}. Iterations stop once ${\hat{C}_{\left( {i}%
\right) }^{\left( j\right) }}$ is within a specified fraction of its value
at the end of the previous iteration or a specified maximum number is
reached. The fraction should be sufficiently small (perhaps 10\%) that the
performance loss will be insignificant.

\section{BLIND CSI ESTIMATION}

The EM algorithm in Section III generates updated CSI estimates as shown in (%
\ref{EQ:ChatRe})--(\ref{EQ:I0}) \emph{after} the initial coherent
demodulation and decoding of receiver iteration $j=0$. In \cite{MILCOM06}--%
\cite{MILCOM07}, the initial CSI estimates were obtained with the aid of
pilot symbols. In this section, two methods for blind estimation of the
initial CSI parameters $\bm{\hat{\theta}}_{(i_{max})}^{(0)}=\left( {\hat {C}%
_{(i_{max})}^{(0)},\hat{I}_{0(i_{max})}^{(0)}}\right) $ are presented, with
the special case of perfect phase information at the receiver examined
first.

\subsection{Perfect Phase Information at Receiver}

The carrier synchronization provided by a phase-locked loop in several
second and third-generation cellular standards such as IS-95 and CDMA20001x
can be exploited to obviate the need to estimate the channel phase (which is
also potentially provided by 2\% piloting \cite{Pados07}). Assuming perfect
phase information at the receiver, the fading amplitude is real-valued and
nonnegative, and (\ref{EQ:ChatIm}) does not have to be computed. A simple
heuristic estimate (denoted as \textit{blind method I}) of $\left( {\hat{C}%
_{(i_{max})}^{(0)},\hat{I}_{0(i_{max})}^{(0)}}\right) $ for each fading
block can be obtained from the received symbols as
\begin{equation}
{{\hat{C}_{(i_{max})}^{\left( {0}\right) }}}=\frac{2}{{n_{FB}}}%
\sum\limits_{k=1}^{{n_{FB}/2}}{\left\vert {y(k)}\right\vert }
\label{EQ:RoughChat}
\end{equation}%
\begin{equation}
\hat{I}_{0,_{(i_{max})}}^{(0)}=\max \left[ {{D-{\left( \hat{C}%
_{(i_{max})}^{\left( {0}\right) }\right) }^{2}},h\cdot {\left( {\hat{C}%
_{(i_{max})}^{\left( {0}\right) }}\right) }^{2}}\right]  \label{EQ:RoughI0}
\end{equation}%
where
\begin{equation}
D=\frac{2}{{n_{FB}}}{\sum\nolimits_{k=1}^{{n_{FB}/2}}\left\vert {y(k)}%
\right\vert ^{2}}  \label{EQ:RoughD}
\end{equation}%
represents the average power of the received symbols, and ${{D-{\left( \hat{C%
}_{(i_{max})}^{\left( {0}\right) }\right) }^{2}}}$ is the difference between
that power and the estimated average power of a desired symbol. Equation (\ref%
{EQ:RoughChat}) would provide a perfect estimate in the absence of noise and
interference. The parameter $h>0$ is chosen such that ${\left( {\hat{C}%
_{(i_{max})}^{(0)}}\right) ^{2}%
\mathord{\left/ {\vphantom {{\hat
C_{(i_{max})}^{(0)} } {\hat I_{0}^{(0)} }}} \right.
\kern-\nulldelimiterspace} {\hat I_{0,(i_{max})}^{(0)} }}$ does not exceed
some maximum value. Ideally, $h$ is a function of $E_{s}/N_{0}$, but in this
paper a constant $h=0.1$ is always used for simplicity.

\subsection{Complexity Analysis}

Although the EM estimation is a relatively low-complexity iterative approach
to maximum-likelihood estimation, it consumes a much larger number of
floating-point operations than pilot-assisted schemes do. To evaluate the
complexity of the EM estimator in terms of required real additions and
multiplications per block of $N_{1}$ code symbols, each complex addition is
equated to two real additions, each complex multiplication is equated to
four real multiplications, and divisions are equated with multiplications.
Equations (\ref{EQ:ChatRe})$-$(\ref{EQ:I0}) require $j_{\max}i_{\max}\left(
6N_{1}+4\right) $ real additions and $j_{\max}i_{\max}\left(
12N_{1}+4\right) $ real multiplications. Equations (18) and (19) require $%
6j_{\max}i_{\max}$ real additions, $30j_{\max}i_{\max}$ real
multiplications, and the computation of $4$ exponentials. The initial
estimates calculated using (\ref{EQ:RoughChat})$-$(\ref{EQ:RoughD}), which
only need to be computed once prior to the first EM iterations, require 2$%
N_{1}$ real additions, $8N_{1}+7$ real multiplications, and the computation
of the maximum of two real numbers. A PACE receiver that uses only pilot
symbols for CSI estimation requires $6N_{1}+4$ real multiplications and $%
12N_{1}+4$ real multiplications to compute (\ref{EQ:ChatRe})$-$(\ref{EQ:I0})
once and does not need to compute the other equations. Thus, EM estimation
increases the amount of computation for CSI estimation by a factor of more
than $j_{\max}i_{\max}$\ relative to PACE.

\subsection{No Phase Information at Receiver}

The initial CSI estimates proposed in (\ref{EQ:RoughChat}) and (\ref%
{EQ:RoughI0}) for blind method I are expected to be degraded significantly
when the phase information is also unknown, since an arbitrary initial phase
value (e.g., 0 radians) must be assumed. To circumvent this problem, the
initial receiver iteration consists of hard-decision demodulation and
channel decoding, after which each decoded bit is used as $\bar {x}%
_{(i_{max})}^{(0)}(k)$ in (\ref{EQ:ChatRe})--(\ref{EQ:I0}). This step is
followed by the regular EM estimation process in subsequent receiver
iterations. This approach for the initial CSI estimates, which is referred
to as \textit{blind method II} in the sequel, results in increased receiver
latency relative to the previous method when phase information is not available.

\subsection{Blind-PACE Estimation Tradeoffs}

The previously proposed iterative DS-CDMA receiver with PACE \cite{MILCOM06}%
--\cite{MILCOM07} is considered as the benchmark for comparison with the
proposed receiver. Assuming an identical transmit-power constraint and
information bit-rate in both cases, the elimination of pilots creates the
following possibilities for methods I and II:

\begin{itemize}
\item (Case $A$) An increase in the number of transmitted information
symbols.

\item (Case $B$) An increase in transmitted information-symbol duration.

\item (Case $C$) An increase in the number of transmitted parity symbols
(lowered IRA code rate).
\end{itemize}

The modifications listed above offset the loss in system performance due to
the degraded CSI estimation obtained from blind methods I and II with
respect to PACE. The no-pilot cases $A$, $B$, and $C$ have the same
transmitted frame duration as the frame with pilot symbols. Cases $A$, $B$,
and $C$ provide the most favorable throughput, spectral efficiency, and bit
error rate, respectively. Numerical evaluations of each of these cases are
presented in the next section. Although a correlated fading model is assumed
in the simulations, no filtering is used to exploit this correlation in
order to maintain the robustness of the proposed estimator.

\section{SIMULATION RESULTS}

In all the simulations, the block sizes are equal, and the information-bit
rate is 100 kb/s. Increasing the block sizes increases the accuracy of the
EM estimators, but decreasing the block sizes allows closer tracking of the
channel parameters and includes more diversity in the receiver computations.
In most of the simulations, except where stated, we set $n_{FI}$=$n_{FB}=40$
and spreading factor $g=31$. The number of closed-loop receiver iterations
is set to $j_{max}=9$, as there is insignificant performance improvement for
$j_{max}>9$. The number of internal EM iterations is $i_{\max }=10.$ There is one decoder iteration per receiver iteration.
A IRA code (data block size $K$ = 1000) with sum-product
algorithm decoding \cite{Proakis} is used without channel interleaving. The IRA code is rate-$1/2$ when PACE is used.
Jakes correlated fading of the desired signal and a mobile velocity of 120
km/hr are assumed. Flat fading is assumed in most of the simulations,
whereas a frequency-selective channel is examined in Section~\ref{sec:RAKE}.
The iterative PACE receiver considered for comparison contains 9.1\%
pilot-symbol overhead, which has been shown to have a decoding performance
close to the conventional 3GPP LTE receiver \cite{MILCOM07}. For each
scenario tested, 5000 Monte Carlo simulation trials were conducted. To avoid
repetition, a selection of representative examples out of the many possible
combinations of channel coding, phase information, interference models, and
no-pilot modifications are presented next.

The bit error rate (BER) is calculated as a function of $E_{b}/N_{0},$ where
$E_{b}=(N/2K)E_{s}$ is the energy per bit. The information throughput is a
vital performance criterion in addition to the BER. One of the primary
motivations in removing pilot symbols is the expectation of achieving
greater throughput, even though the BER performance may be degraded
marginally. We define throughput $R$ as
\begin{equation}
R=\frac{{{\text{information bits in a codeword}}}}{{{\text{codeword duration}%
}}}\times\left( {1-BER}\right) \hspace{0.1in}\text{bits/s}.
\end{equation}

\subsection{Single-user environment, perfect phase knowledge}
\begin{figure}[pth]
\centering
\includegraphics[width=\linewidth]{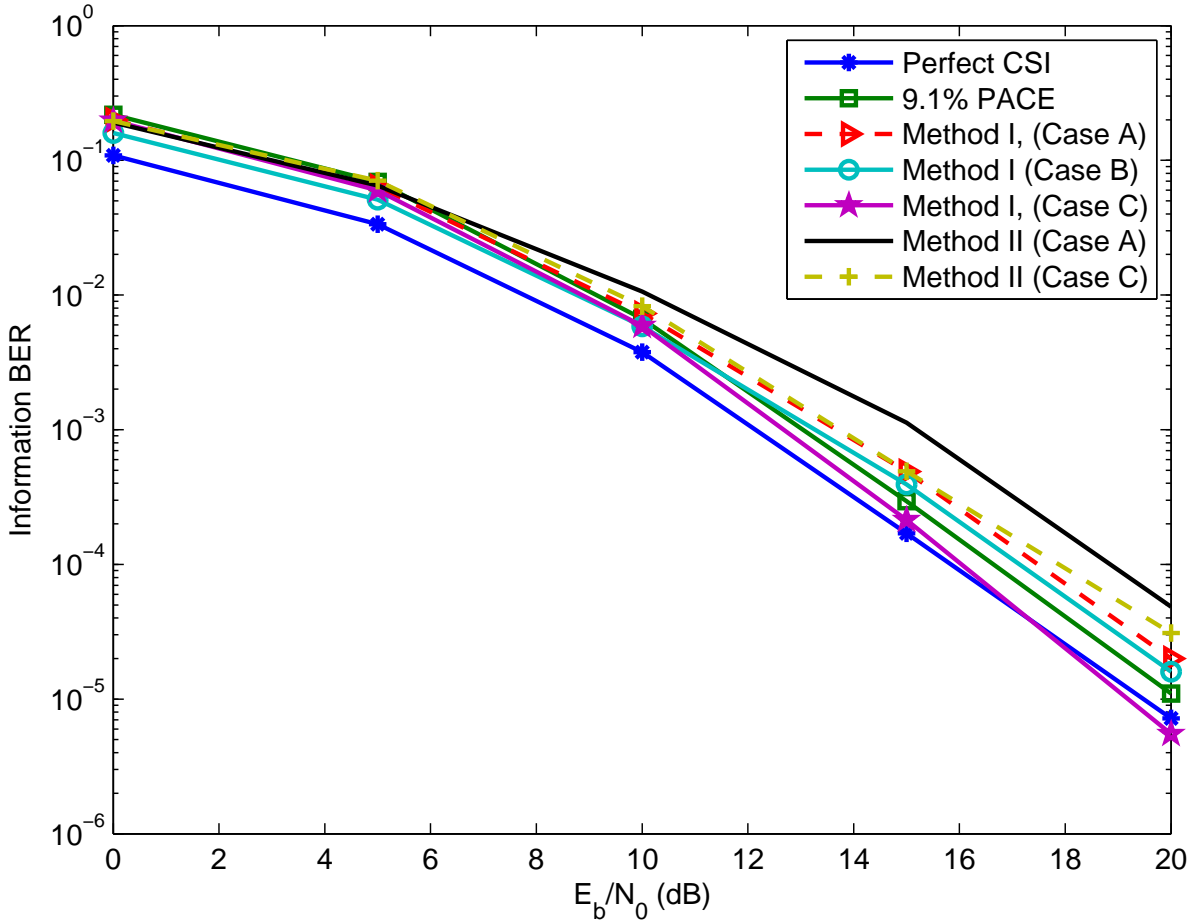}
\caption{BER versus $E_{b}/N_{0}$ for IRA-coded iterative receiver in
single-user environment with phase provided by PLL.}
\label{fig:LDPC_SU_BER1}
\end{figure}
\begin{figure}[pth]
\centering
\includegraphics[width=\linewidth]{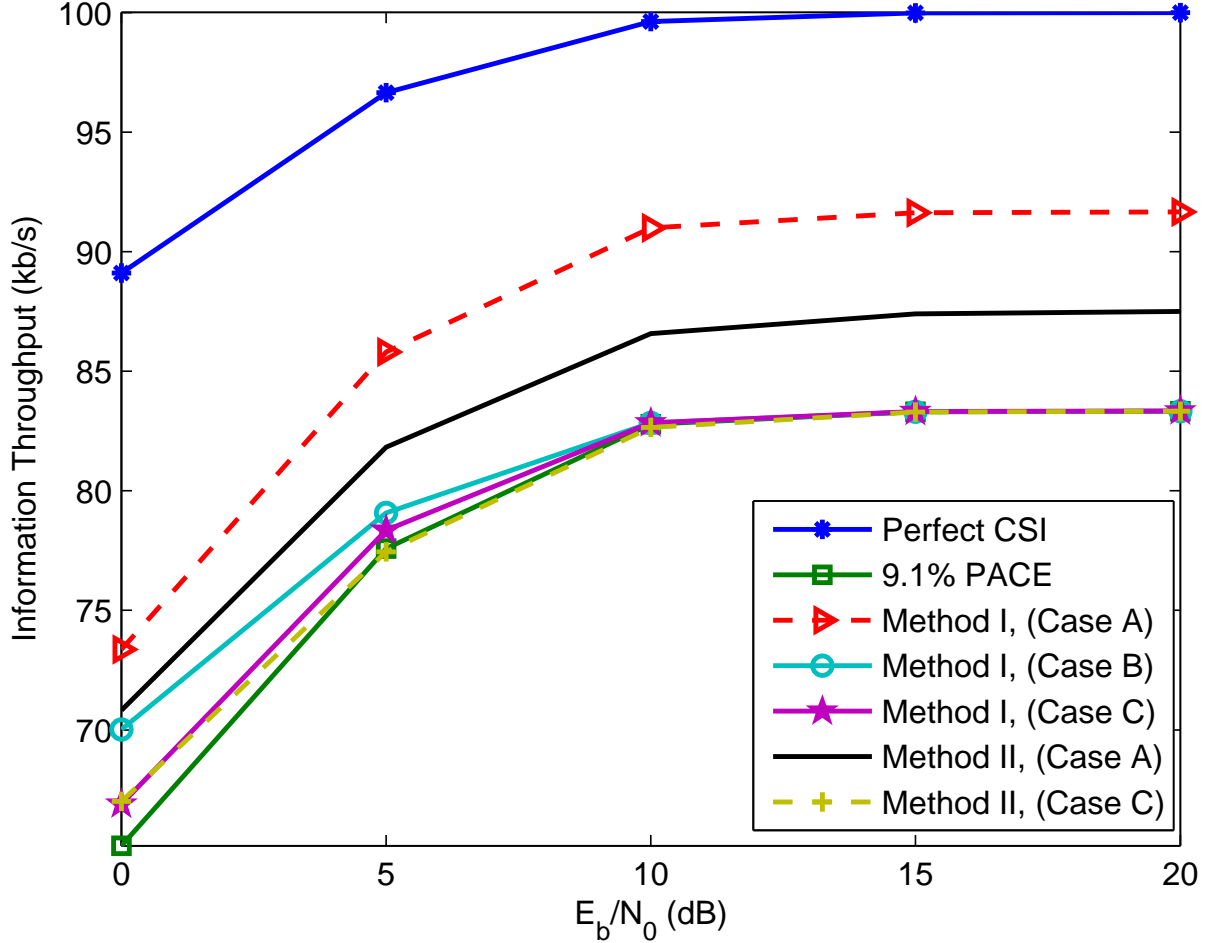}
\caption{Information throughput versus $E_{b}/N_{0}$ for IRA-coded iterative
receiver in single-user environment with phase provided by PLL.}
\label{fig:IRA_SU_Thru1}
\end{figure}
For the first set of results in Figs. \ref{fig:LDPC_SU_BER1}--\ref%
{fig:IRA_SU_Thru1}, a single-user environment and perfect phase knowledge at
the receiver are assumed. Fig. \ref{fig:LDPC_SU_BER1} displays the BER
versus $E_{b}/N_{0}$ for an IRA-coded iterative receiver operating with
perfect CSI, PACE, blind method I with cases $A$, $B$, and $C$, and blind
method II with cases $A$ and $C$, respectively. The key observation is that
blind method II is worse than method I by 2 dB at $BER=10^{-3}$ for both
case $A$ and case $C$, which illustrates the well-known sensitivity of the
EM algorithm to the accuracy of the initial estimates.

The addition of extra parity bits to blind method I (case $C$,
rate-1000/2200) offers the greatest improvement in BER, surpassing even the
rate-1/2 code with perfect CSI at high $E_{b}/N_{0}$. The increase in number
of information symbols (case $A$) results in the worst BER performance with
a separation of 1 dB and 0.5 dB from PACE and case $B$ at $BER=10^{-3}$,
respectively.

The various scenarios featured in Fig. \ref{fig:LDPC_SU_BER1} were also
tested under a slow-fading channel with mobile velocity of 10 km/hr. It was
observed that all the BER curves were shifted towards the right by up to 7
dB at $BER=10^{-3}$, but the overall trends among the different cases
remained the same.

Fig. \ref{fig:IRA_SU_Thru1} exhibits information throughput $R$ versus $%
E_{b}/N_{0}$ for the IRA-coded iterative receiver with the scenarios of Fig.~%
\ref{fig:LDPC_SU_BER1}. The throughput advantage of case $A$ is achieved
even though no pilot symbols are used at all; i.e., the initial estimation
is totally blind. It is evident that increasing the symbol duration or
adding additional parity information does not give the proposed blind
methods any significant advantage in throughput over PACE. Both blind
methods with cases $B,C$ and PACE provide about 20\% less throughput than
the receiver with perfect CSI.

\subsection{Multiuser environment, unknown phase}

A 4-user interference environment with equal mean bit energies for all users
at the receiver, $E_{b}/N_{0}=20$ dB, and no phase information at the
receiver is examined next. It is assumed that both the interference levels
and the unknown phase are constant during each subframe. Each interference
signal experiences independent Jakes correlated fading and uses independent
data and Gold sequences with respect to the desired signal. The simulation
uses chip-synchronous interference signals, which is a pessimistic
worst-case assumption \cite{Torrieri}. Two variations of CSI estimation are
examined here: \emph{partially adaptive} with only fading coefficient $\hat{C%
}_{(i)}^{(j)}$ being estimated using (\ref{EQ:ChatRe}), (\ref{EQ:ChatIm}),
and $\hat{I}_{0(i)}^{(j)}$ set equal to $N_{0}$ for all subframes; and \emph{%
fully adaptive} estimation of both $\hat{C}_{(i)}^{(j)}$ and $\hat{I}%
_{0(i)}^{(j)}$ using (\ref{EQ:ChatRe}), (\ref{EQ:ChatIm}), and (\ref{EQ:I0}).

\begin{figure}[pth]
\centering
\includegraphics[width=\linewidth]{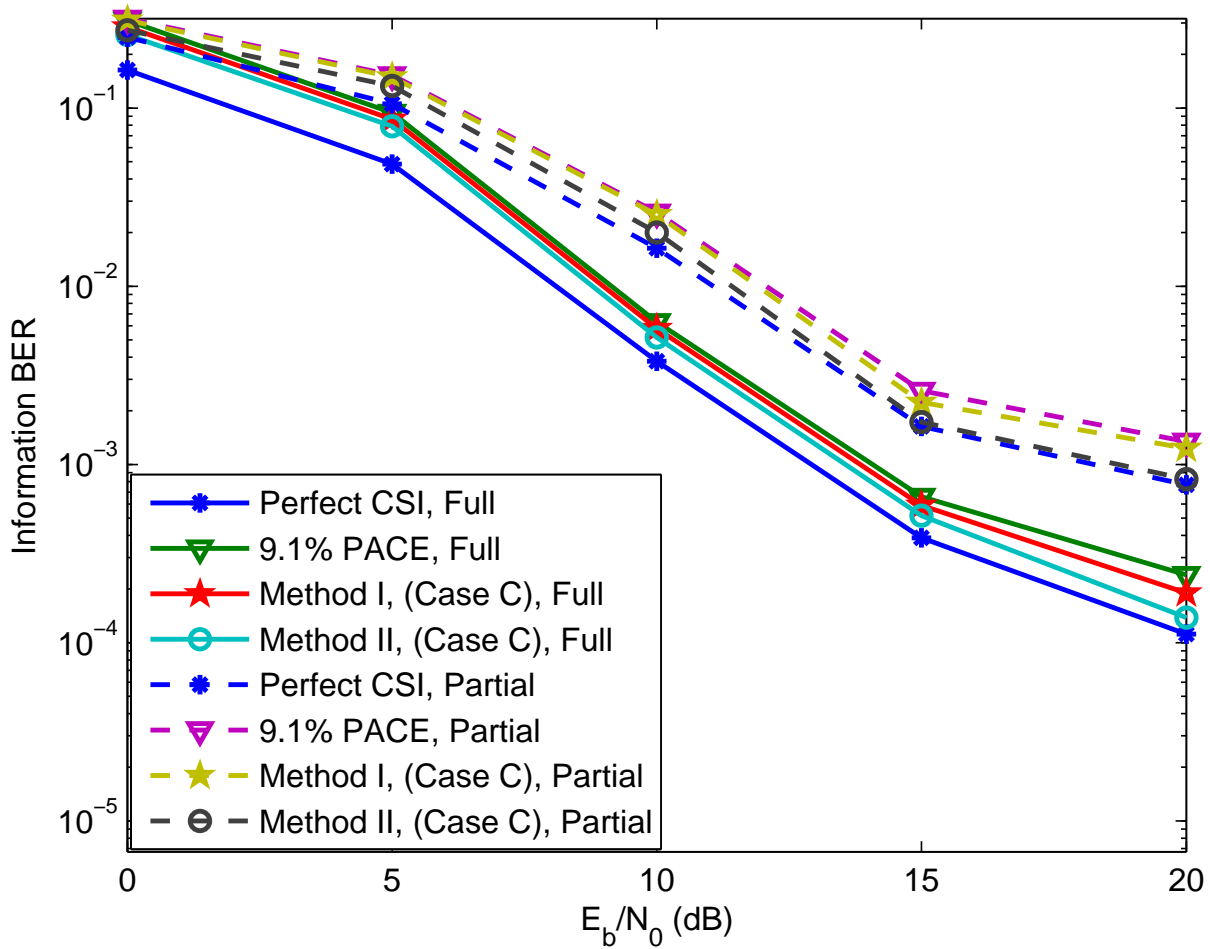}
\caption{BER versus $E_{b}/N_{0}$ for IRA-coded iterative receiver affected
by MAI from 4 users, fully and partially adaptive estimation, and unknown
phase.}
\label{fig:IRA_MUI_BER}
\end{figure}

Fig. \ref{fig:IRA_MUI_BER} displays IRA-coded BER versus $E_{b}/N_{0}$ for
partially and fully adaptive CSI estimation per fading block and case $C$
for both blind methods. The mismatch of $\hat{I}_{0}$ and the true value of $%
I_{0}$ at the demodulator and decoder results in a high error floor for the
partially adaptive cases. The intuition behind the error floor is that the
partially adaptive estimator overestimates the true
signal-to-interference-plus-noise ratio (SINR) by disregarding the MAI, with
the degree of overestimation increasing with SINR. For IRA codes, it was
shown in \cite{Cheun05} that both under- and overestimation of the SINR
degrades the IRA decoder performance. The fully adaptive estimation offers a
more accurate SINR estimate and, hence, suppresses interference and reduces
the error floor significantly. This interference suppression is achieved
without using the far more elaborate multiuser and signal cancellation
methods that could be implemented in a DS-CDMA receiver. For both partially
and fully adaptive estimation, it is observed that blind method II now
outperforms method I due to better phase estimation, whereas both blind
methods outperform PACE at $BER=10^{-3}$ due to the added parity
information.

\begin{figure}[pth]
\centering
\includegraphics[width=\linewidth]{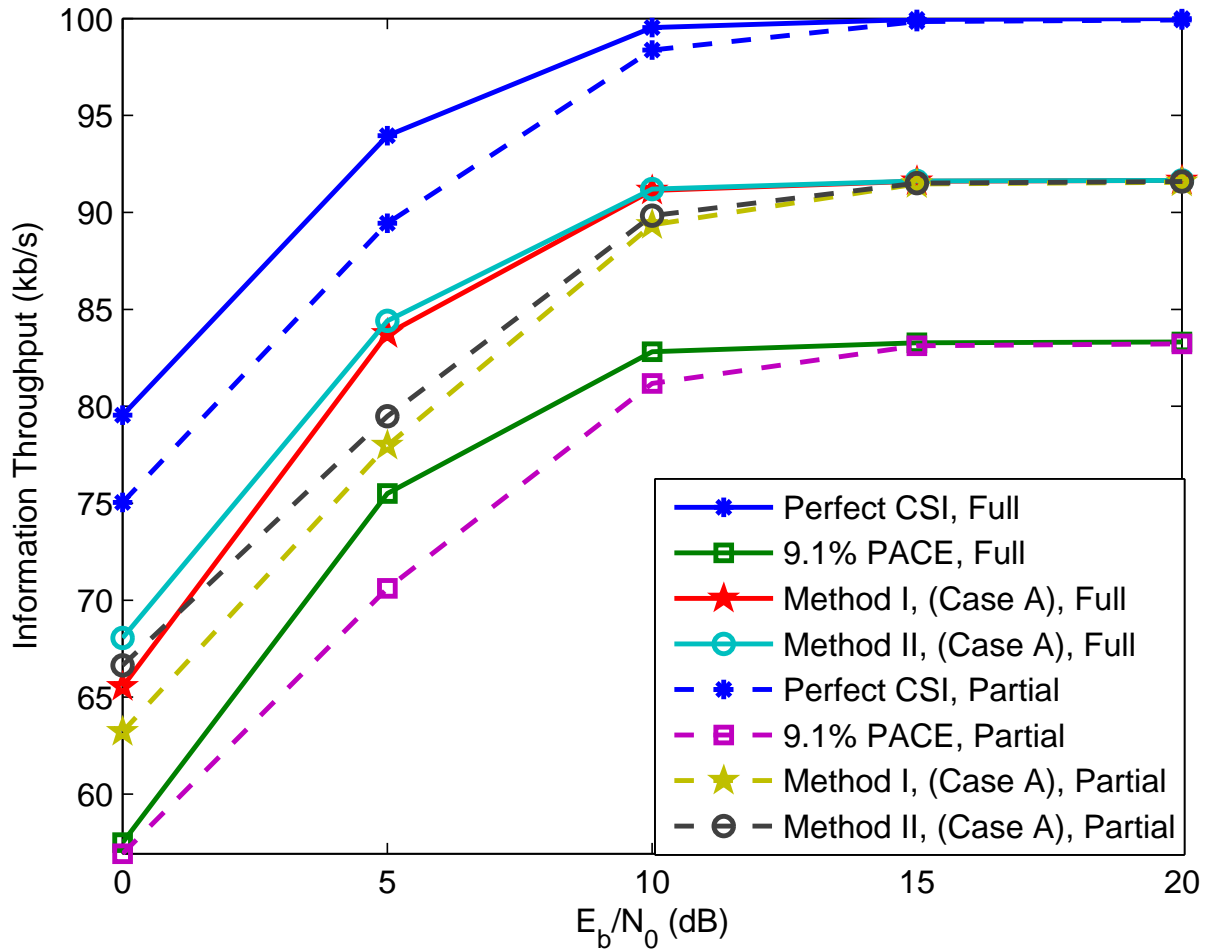}
\caption{Information throughput versus $E_{b}/N_{0}$ for IRA-coded iterative
receiver affected by MAI from 4 users, fully and partially adaptive
estimation, and unknown phase.}
\label{fig:IRA_MUI_Thru}
\end{figure}
Fig. \ref{fig:IRA_MUI_Thru} demonstrates the IRA-coded receiver throughput
offered by the proposed methods under MAI from 4 users.
The blind methods always provide a better throughput compared with PACE; for
example, method I with case $A$ is superior by 9\% to both PACE scenarios
when $E_{b}/N_{0}>5$ dB. It is observed that both partial and fully-adaptive
estimation methods offer a similar asymptotic throughput, which indicates
that partial CSI estimation may be sufficient for applications with a
non-stringent BER criterion. On the other hand, error-critical
applications requiring less than $BER=10^{-3}$ must use the fully adaptive
CSI estimation, as seen from Fig. \ref{fig:IRA_MUI_BER}.

\subsection{Varying fading-block size, unknown phase}

In urban mobile environments, the phase can be expected to change
significantly after approximately $\frac{0.01}{f_{d}}$ s to $\frac{0.04}{%
f_{d}}$ s, where $f_{d}$ is the maximum Doppler shift. For the assumed
mobile velocity of 120 km/hr, this time range corresponds to roughly 10 to
40 code bits at 100 kb/s. The fading and interference block sizes $n_{FB}=$ $%
n_{IB}$ are therefore varied accordingly, and \emph{no} phase information is
assumed to be available at the receiver for the next set of results.

\begin{figure}[pth]
\centering
\includegraphics[width=\linewidth]{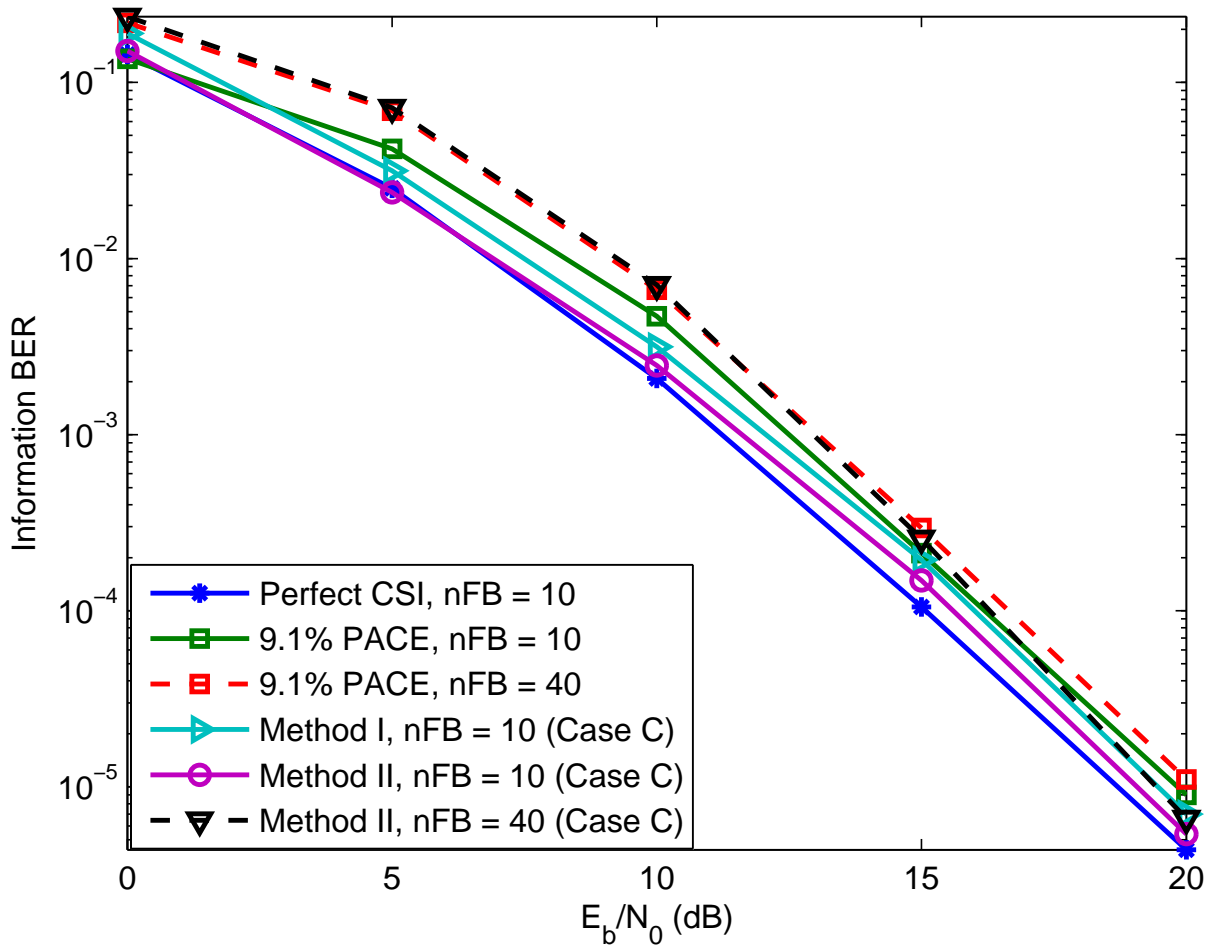}
\caption{BER versus $E_{b}/N_{0}$ for IRA-coded iterative receiver in
single-user environment, varying $n_{FB}$, and unknown phase.}
\label{fig:IRA_SU_noPLL_BER_nFB}
\end{figure}
Fig. \ref{fig:IRA_SU_noPLL_BER_nFB} displays fully adaptive IRA-coded BER
versus $E_{b}/N_{0}$ for blind methods I and II with case $C$, 9.1 \% PACE,
and perfect CSI decoding for $n_{FB}=10$ and $40$ in a single-user
environment. An improvement of 1 to 2 dB was observed for all methods for
the smaller fading-block size of $n_{FB}=10$ due to the increased fading
diversity. The throughput with case $A$ is shown in Fig. \ref%
{fig:IRA_SU_noPLL_Thru_nFB}. It is observed that the throughput gains of the
proposed blind methods over PACE (roughly 9\% at intermediate to high $%
E_{b}/N_{0}$) are preserved even when the phase is initially unknown at the
receiver.
\begin{figure}[pth]
\centering
\includegraphics[width=\linewidth]{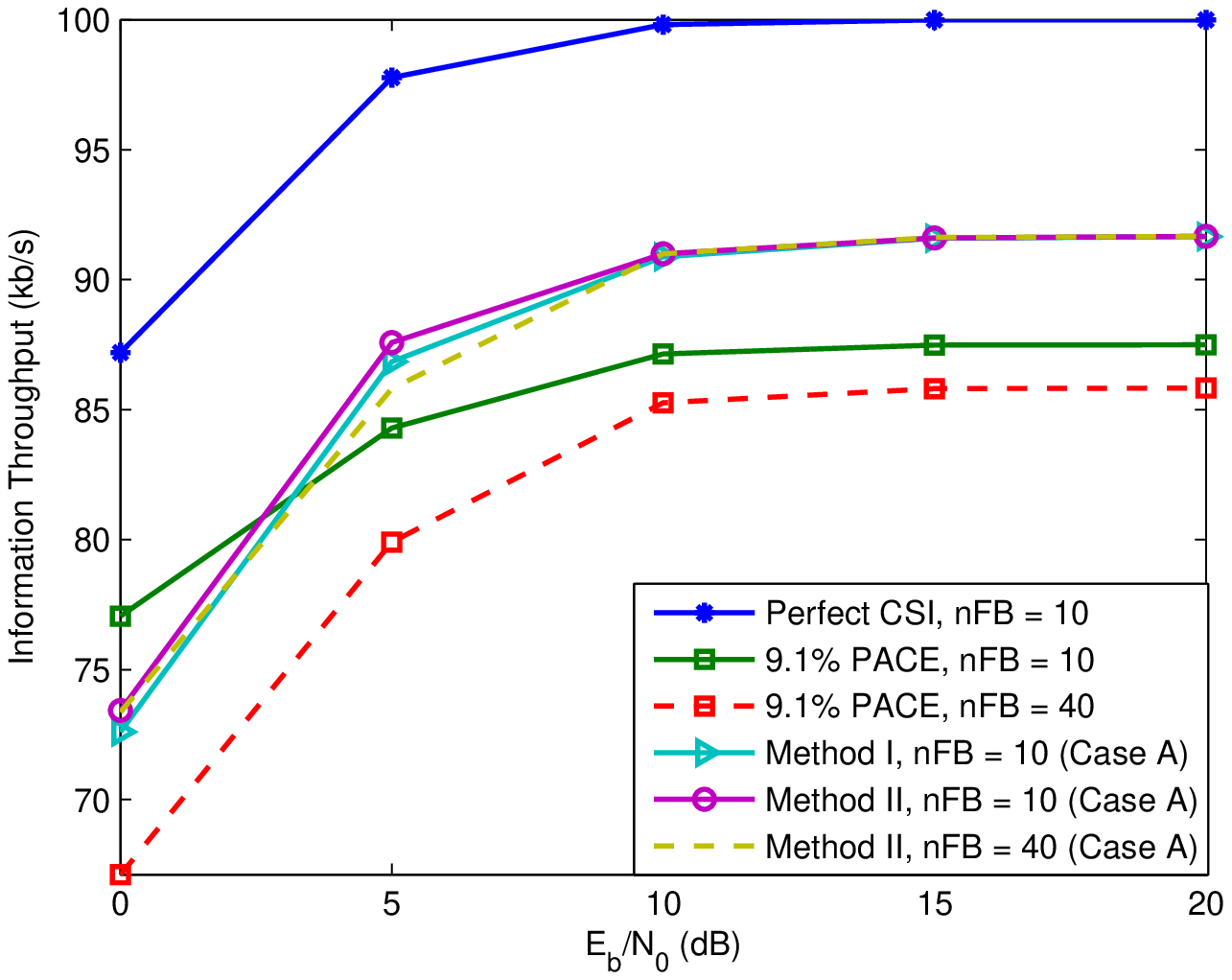}
\caption{Information throughput versus $E_{b}/N_{0}$ for IRA-coded iterative
receiver in single-user environment, varying $n_{FB}$, and unknown phase.}
\label{fig:IRA_SU_noPLL_Thru_nFB}
\end{figure}
\subsection{Varying MAI, unknown phase}

\begin{figure}[pth]
\centering
\includegraphics[width=\linewidth]{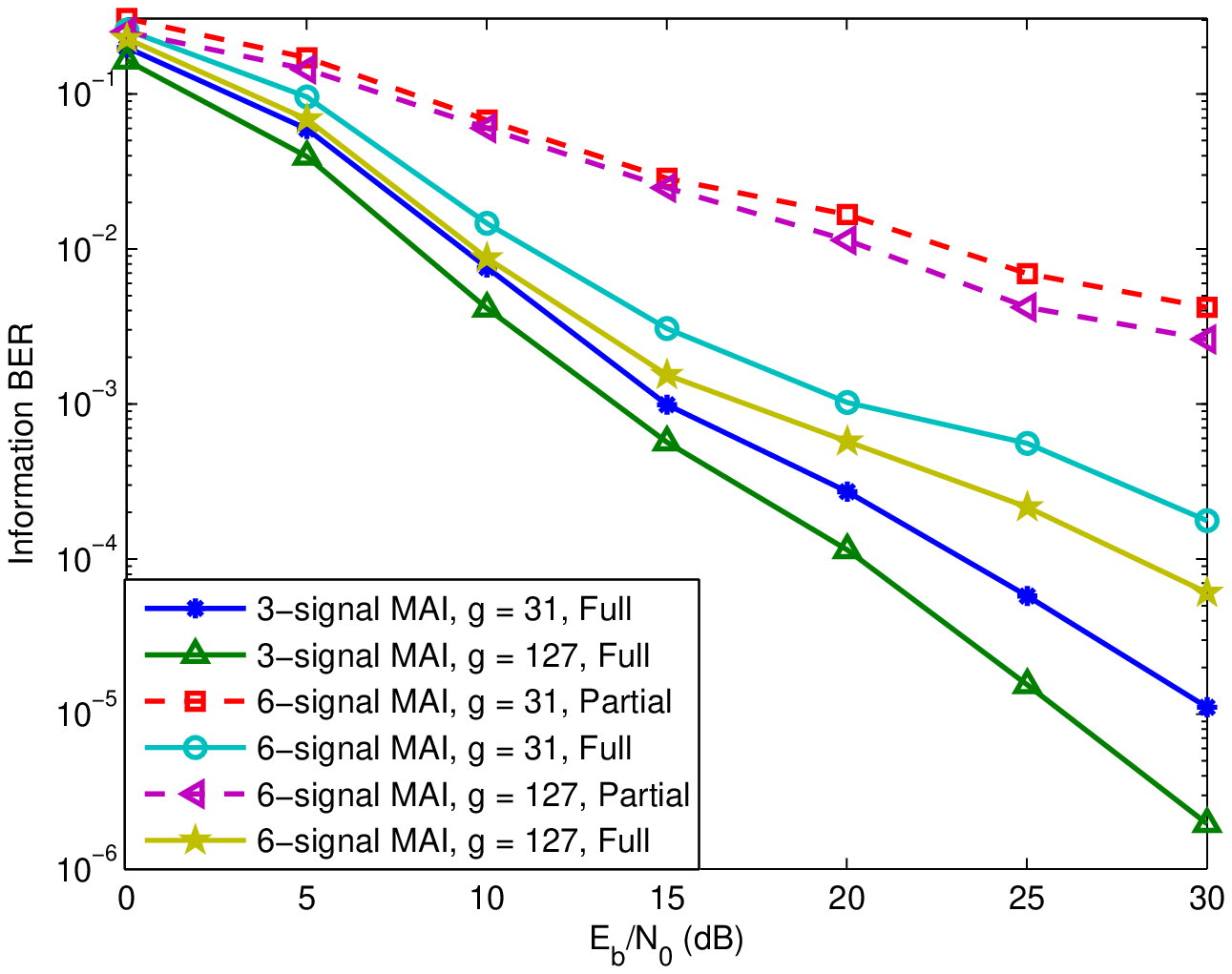}
\caption{BER versus $E_{b}/N_{0}$ for IRA-coded iterative receiver affected
by an unknown phase and various spreading factors, MAI levels, and degrees
of adaptation.}
\label{fig:Nov3_dsl_SF}
\end{figure}
IRA-coded iterative receiver performance with blind method II, case $C$ is
examined for 3 and 6 MAI signals with equal mean bit energies for all users
at the receiver in Fig. \ref{fig:Nov3_dsl_SF}. The partially adaptive
estimation is unable to cope with the interference caused by 6 MAI signals
regardless of the spreading factor, whereas the fully adaptive estimation
offers a substantial improvement in BER. The benefit of an increased
spreading factor ($g=127$ versus $g=31$) is more apparent at low bit error
rates for fully adaptive estimation. For example, the fully adaptive
estimation with 3 MAI signals improves by a factor of approximately 5 dB at $%
BER=10^{-5}$, despite nonorthogonal spreading sequences and imperfect CSI.

\subsection{Multipath channel}\label{sec:RAKE}
A DS-CDMA system can exploit a frequency-selective fading
channel by using a Rake receiver. As an example, we assume a channel with
three resolvable multipath components (with known delays) of the desired
signal and a Rake combiner with three corresponding fingers. The multipath
components undergo independent fading across the fingers, but follow the
Jakes correlated fading assumption over time. The multipath components follow an
exponentially decaying power profile across the fingers, i.e., $E\left[ {%
\alpha _{l}}\right] ^{2}=e^{-(l-1)}$, $l=1,2,3$. Each interference signal
has the same power level in each finger and undergoes independent Jakes
correlated fading. The assumption of independent multipath fading amplitude
and phase coefficients for the desired signal allows us to apply the
proposed EM-based channel estimation scheme separately in each finger. The
Rake combiner performs maximal-ratio combining (MRC) of the received symbol
copies based on channel and interference-PSD estimates computed at all
fingers. The MRC decision statistic obtained from the Rake combiner is then
passed to the QPSK demodulator metric generator, which generates soft inputs
for the common channel decoder. The channel decoder soft outputs are fed
back to the three channel estimator blocks, which then recompute updated
channel coefficients, as described in Section~III.

\begin{figure}[tph]
\centering
\includegraphics[width=\linewidth]{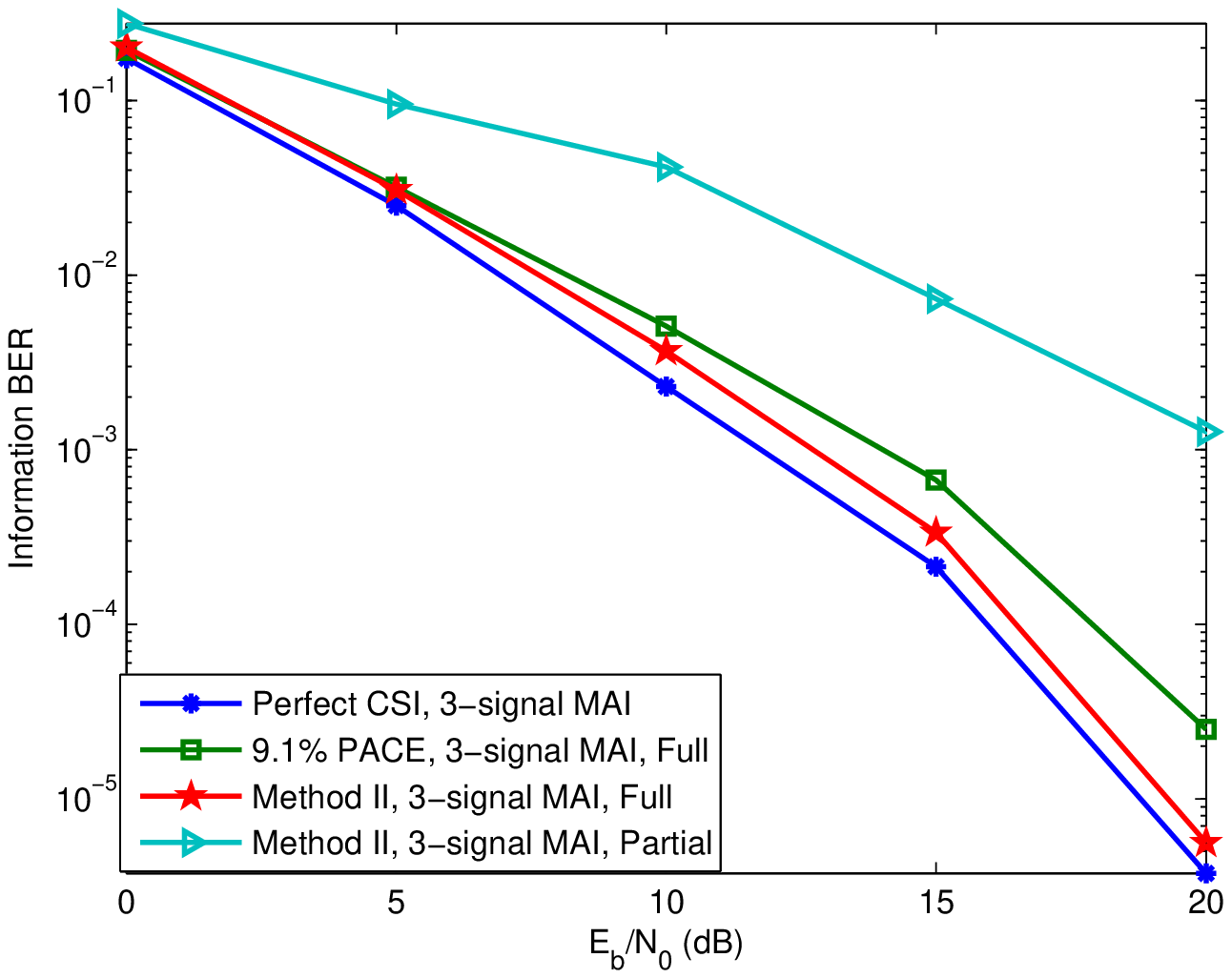}
\caption{BER versus $E_{b}/N_{0}$ for IRA-coded iterative RAKE receiver with
three resolvable multipath components and three fingers.}
\label{fig:RAKE}
\end{figure}
Fig.~\ref{fig:RAKE} displays the Rake receiver performance for various
levels of MAI with Method II under case $C$, where all users have length-127
Gold sequences. It is observed that the additional diversity due to Rake
combining improves performance as expected, but the performance disparity
between partially and fully adaptive estimation remains large.

\section{Conclusions}

It has been shown that pilot symbols are not essential to the effectiveness
of DS-CDMA receivers with coding, coherent detection, and channel
estimation. If the pilot symbols are replaced by information symbols, the
throughput increases relative to PACE whether or not interference is
present. If the BER is the primary performance criterion, then replacing the
pilot symbols by parity symbols gives a lower BER than PACE. If the spectral
efficiency is of primary importance, then extending the symbol duration
after the removal of the pilot symbols offers an improvement relative to
PACE, albeit at the cost of a slight increase in the BER.

The estimation of the interference PSD has been shown to enable the
significant suppression of interference. This suppression is achieved
without using the far more elaborate multiuser and signal cancellation
methods that could be implemented in a DS-CDMA receiver.

\section*{Acknowledgment}
The second author would like to thank Avinash Mathur for his assistance in the early stages of this work.


\begin{thebibliography}{99}
\bibitem{3GPP} 3GPP \emph{Physical Channels and Modulation} [Online].
Available: http://www.3gpp.org/ftp/Specs/archive/36\%5Fseries/36.211/

\bibitem{Cavers} J. K. Cavers, ``An analysis of pilot symbol assisted
modulation for Rayleigh fading channels," \emph{IEEE Trans. Veh. Tech}.,
vol. 40, pp. 686 - 693, Nov. 1991.

\bibitem{Globe99} P. Hoeher and F. Tufvesson, ``Channel estimation with
superimposed pilot sequence," in \emph{Proc. IEEE GLOBECOM}, pp. 2162 -
2166, 1999.

\bibitem{Blind94} L. Tong, G. Xu, and T. Kailath, ``Blind identification and
equalization based on second-order statistics: A time domain approach,"
\emph{IEEE Trans. Inf. Theory}, vol. 40, pp. 340 - 349, Mar. 1994.

\bibitem{Poor98} X. Wang and H. V. Poor, ``Blind adaptive multiuser
detection in multipath CDMA channels based on subspace tracking," \emph{IEEE
Trans. Signal Process}., vol. 46, pp. 3030 - 3044, Nov. 1998.

\bibitem{SNR} T. A. Summers and S. G. Wilson, ``SNR mismatch and online
estimation in turbo decoding," \emph{IEEE Trans. Commun}., vol. 46, pp. 421
- 423, Apr. 1998.

\bibitem{Mackay_noise} D. J. MacKay and C. P. Hesketh, ``Performance of low
density parity check codes as a function of actual and assumed noise
levels," \emph{Electr. Notes Theor. Comput. Sci.}, vol. 74, pp. 89 - 96,
2003.

\bibitem{EM} G. J. McLachlan and T. Krishnan, \emph{The EM Algorithm and
Extensions}, Wiley, 1997.

\bibitem{EM_1977} A. P. Dempster, N. M. Laird, and D. B. Rubin,
``Maximum-likelihood from incomplete data via the EM algorithm," \emph{J.
Roy. Stat. Soc}., vol. 39, pp. 1 - 38, 1977.

\bibitem{EM_99} H. Jafarian and S. Pasupathy, ``EM-based recursive
estimation of channel parameters," \emph{IEEE Trans. Commun}., vol. 47, pp.
1297 - 1302, Sep. 1999.

\bibitem{EM_Wang1} B. Lu, X. Wang, and K. R. Narayanan, ``LDPC-based
space-time coded OFDM systems over correlated fading channels: Performance
analysis and receiver design," \emph{IEEE Trans. Commun}., vol. 50, pp. 74 -
88, Jan. 2002.

\bibitem{EM_Wang2} B. Lu, X. Wang, and Y. Li, ``Iterative receivers for
space--time block coded OFDM systems in dispersive fading channels," \emph{%
IEEE Trans. Wireless Commun.}, vol. 1, pp. 213 - 225, Apr. 2002.

\bibitem{EM_ref} A. Kocian and B. H. Fleury, ``EM-based joint data detection
and channel estimation of DS-CDMA signals," \emph{IEEE Trans. Commun}., vol.
51, pp. 1709 - 1720, Oct. 2003.

\bibitem{Poor96} L. B. Nelson and H. V. Poor, ``Iterative multiuser
receivers for CDMA channels: An EM-based approach," \emph{IEEE Trans. Commun}%
., vol. 44, pp. 1700 - 1710, Dec. 1996.

\bibitem{Caire2001} M. Kobayashi, J. Boutros, and G. Caire, ``Successive
interference cancellation with SISO decoding and EM channel estimation,"
\emph{IEEE J. Selec. Areas Commun}., vol. 19, pp. 1450 - 1460, Aug. 2001.

\bibitem{EM_Mitra} S. Wu, U. Mitra, and C.-C. Jay Kuo, ``Iterative joint
channel estimation and multiuser detection for DS-CDMA in
frequency-selective fading channels," \emph{IEEE Trans. Signal Process}.,
vol. 56, pp. 3261 - 3277, Jul. 2008.

\bibitem{ChengTrans07} S. Cheng, M. C. Valenti, and D. Torrieri, ``Robust
iterative noncoherent reception of coded FSK over block fading channels,"
\emph{IEEE Trans. Wireless Commun}., vol. 6, pp. 3142 - 3147, Sep. 2007.

\bibitem{Proakis} J. G. Proakis and M. Salehi, \emph{Digital Communications,
$5^{th}$ Ed}. New York: McGraw-Hill, 2008.

\bibitem{MILCOM06} D. Torrieri, E. Ustunel, H. M. Kwon, S. Min, and D. H.
Kang, ``Iterative CDMA receiver with EM channel estimation and turbo
decoding," in \emph{Proc. IEEE MILCOM}, Washington D.C., pp. 1 - 6, Oct.
2006.

\bibitem{VTC07} D. Torrieri, A. Mathur, A. Mukherjee, and H. M. Kwon,
``Iterative LDPC CDMA receiver under time-varying interference," in \emph{%
Proc. 65th IEEE Veh. Tech. Conf.}, Dublin, Ireland, pp. 1986 - 1989, Apr.
2007.

\bibitem{MILCOM07} D. Torrieri, A. Mukherjee and H. M. Kwon, ``Iterative EM
channel estimation for DS-CDMA receiver using LDPC codes with M-ary
modulation," in \emph{Proc. IEEE MILCOM}, Orlando, Fl, pp. 1 - 6, Oct. 2007.

\bibitem{Vandendorpe2007} X. Wautelet, C. Herzet, A. Dejonghe, J. Louveaux,
and L. Vandendorpe, ``Comparison of EM-based algorithms for MIMO channel
estimation," \emph{IEEE Trans. Commun.}, vol. 55, no. 1, pp. 216 - 226, Jan.
2007.

\bibitem{Choi} J. Choi, ``An EM-based iterative receiver for MIMO-OFDM under
interference-limited environments," \emph{IEEE Trans. Wireless Commun}.,
vol. 6, no. 11, pp. 3994 - 4003, Nov. 2007.

\bibitem {Psar1}I. N. Psaromiligkos, S. N. Batalama, and M. J. Medley, ``Rapid
combined synchronization/demodulation structures for DS-CDMA systems - Part I:
algorithmic developments," \emph{IEEE Trans. Commun}., vol. 51, no. 6, pp. 983 - 994,
June 2003.

\bibitem {Psar2}I. N. Psaromiligkos and S. N. Batalama, ``Rapid combined
synchronization/demodulation structures for DS-CDMA systems - Part II: finite
data-record performance analysis," \emph{IEEE Trans. Commun}., vol. 51, no. 7, pp.
1162 - 1172, July 2003.


\bibitem{Stuber} M. C. Necker and G. L. St\"{u}ber, ``Totally blind APP
channel estimation for mobile OFDM systems," \emph{IEEE Trans. Wireless
Commun}., vol. 3, no. 5, pp. 1514 - 1525, Sep. 2004.

\bibitem{Pados07} G. N. Karystinos and D. A. Pados, ``Supervised phase
correction of blind space-time DS-CDMA channel estimates," \emph{IEEE Trans.
Commun}., vol. 55, no. 3, pp. 584 - 592, Mar. 2007.

\bibitem{Yang04} M. Yang, W. E. Ryan, and Y. Li, ``Design of efficiently
encodable moderate-length high-rate irregular LDPC codes," \emph{IEEE Trans.
Commun}., vol. 52, no. 4, pp. 564 - 571, Apr. 2004.

\bibitem{Ryan06} F. Peng, M. Yang, and W. Ryan, ``Simplified eIRA code
design and performance analysis for correlated Rayleigh fading channels,"
\emph{IEEE Trans. Wireless Commun}., vol. 5, no. 4, pp. 720 - 725, Apr. 2006.

\bibitem{Richardson01} T. Richardson, M. A. Shokrohalli, and R. Urbanke,
``Design of capacity approaching irregular low density parity check codes,"
\emph{IEEE Trans. Inf. Theory}, vol. 47, no. 2, pp. 619 - 637, Feb. 2001.

\bibitem{Torrieri} D. Torrieri, \emph{Principles of Spread-Spectrum
Communication Systems}. New York: Springer, 2005.

\bibitem{Valenti} M. C. Valenti and S. Cheng, ``Iterative demodulation and
decoding of turbo coded M-ary noncoherent orthogonal modulation," \emph{IEEE
J. Selected Areas Commun.}, vol. 23, no. 9, pp. 1739 - 1747, Sep. 2005.

\bibitem{Cheun05} W. Oh and K. Cheun, \textquotedblleft Iterative decoding
and channel parameter estimation algorithms for repeat-accumulate codes,"
\emph{IEEE Trans. Commun}., vol. 53, no. 10, pp. 1597 - 1602, Oct. 2005.
\end{thebibliography}
\end{document}